\documentstyle[seceqn,equation]{espart}
\font\st=eufm10 scaled 1200
\font\bi=cmbxti10 scaled 1200
\font\bfs=cmmib10 scaled 1200
\font\bfd=cmbsy10 scaled 1200
\def\bs{\hbox{\bfs\symbol{27}}}
\def\bd{\hbox{\bfd\symbol{1}}}
\begin{document}
\begin{frontmatter}
\title{Groups and nonlinear dynamical systems}
\subtitle{Dynamics on the $SU(2)$ group}
\author{Krzysztof Kowalski and Jakub Rembieli\'nski}
\address{\it Department of Theoretical Physics, University
of \L\'od\'z, ul. Pomorska 149/153,\\ 90-236 \L\'od\'z, Poland}
\journal{Physica D}
\begin{abstract}
An abstract Newton-like equation on a general
Lie algebra is introduced such that orbits of the Lie-group
action are attracting set.  This
equation generates the nonlinear dynamical system satisfied
by the group parameters having an attractor coinciding with
the orbit.  The periodic  solutions of the abstract equation
on a Lie algebra are discussed.  The particular case of the
$SU(2)$ group is investigated.  The resulting nonlinear second-order
dynamical system in ${\bf R}^3$ as well as its constrained
version referring to the generalized spherical pendulum are
shown to exhibit global Hopf bifurcation.
\end{abstract}
\end{frontmatter}

\vspace{1.2cm}

\vspace{.8cm}
\newpage
\section{Introduction}
Over the years, a variety of the group-theoretic methods for
the study of differential equations have been devised [1,2].  It
is scarcely to belabor their importance in the investigation
of integrable systems.  Let us only recall the methods for
constructing group-invariant solutions.  Nevertheless, there
is still a general lack of approaches discussing behaviour
of nonintegrable equations.  Recently, an interesting
approach was proposed by Okni\'nski [3] relying on
the study of nonlinear recurrences by asscociating them with
a discrete-time evolution on a Lie group.  More precisely,
within the Okni\'nski approach the use is made of the
Shimizu-Leutbecher sequence [4,5] which satisfies
the following group recurrence:
\begin{equation}
g_{k+1}=g_khg_k^{-1},
\end{equation}
where $g_k$'s and $h$ are elements of a Lie group $G$.  On
taking a concrete realization of a Lie group $G$ we arrive
at the system of nonlinear difference equations which obey
the group parameters.  For example in the case with the
$SU(2)$ group one obtains the logistic equation satisfied by
one of the coordinates of the system implied by (1.1).  By
choosing $G=E(2)$ and $G=SU(2)$ for the Schimizu-Leutbecher
sequence (1.1) the method was applied to the study of random
walk on the plane [3] and discrete symmetries in
chaos-order transitions [6], respectively.  The
question naturally arises as to whether such approach can be
generalized to involve the case with continuous-time
equations.  An attempt in this direction was made by
Okni\'nski [7] who studied the equation with
delayed argument arising from (1.1) by setting $k$ to be a
continuous parameter.

In this paper we introduce an abstract Newton-like equation
on a general Lie algebra which can be regarded as a
continuous-time algebraic version of (1.1), such that the
orbits of the Lie-group action are attractors.
The corresponding group parameters fixing a
concrete algebra obey the system of nonlinear second-order
differential equations having a limit set coinciding with
the orbit.  In section 2 we introduce the
abstract Newton-like equation on a Lie algebra and study its
asymptotic behaviour.  We show that orbits of the Lie-group
action are attracting set.  Section 3 discusses periodic 
solutions to the abstract Newton-like equation.
Namely, we establish some criteria of the existence of such
solutions.  In section 4 we illustrate the theory by an
example of the $SU(2)$ group.  We first discuss the
nonlinear dynamical system generated by the corresponding
Newton-like equation on the $su(2)$ algebra and its
asymptotic counterpart.  We then study the dynamics on
the orbit i.e. the sphere $S_2$ which is the limit set for
the trajectories of the discussed nonlinear systems.  Such
the dynamics is demonstrated to correspond to a
generalization of the spherical pendulum involving the
presence of the nonpotential forces.  We identify the
periodic solutions and show that the considered dynamical
system on the sphere has the global bifurcation of a limit
cycle from an equilibrium point (Hopf bifurcation).  Using
the fact which is proved in appendix that the sphere $S_2$
is the universal attracting set in the case of the
asymptotic system mentioned above we conclude that this
system has the global Hopf bifurcation.
\section{The Newton-like equation}
In this section we introduce the abstract Newton-like
equation on a general Lie algebra.  By choosing the concrete
algebra this equation generates the nonlinear dynamical
system satisfied by the time-dependent group parameters.
Consider the following second-order differential
equation on a Lie algebra $\hbox{\st\symbol{103}}$:
\begin{equation}
%<2.1>
\mu \ddot X + \nu\dot X +  \rho X + \sigma Y =
e^{iX}Ye^{-iX},\qquad X(0)=X_0,\quad \dot X(0)=\dot X_0,
\end{equation}
where $X(t)$: ${\bf R}\to \hbox{\st\symbol{103}}$ is a
curve in $\hbox{\st\symbol{103}}$,
$Y\in\hbox{\st\symbol{103}}$ is a fixed element, $\mu >
0,\,\nu,\,\rho,\,\sigma\in{\bf R}$ and dot designates
differentiation with respect to time.

Demanding that (2.1) admits the solution on the orbit
\begin{equation}
%<2.2>
{\rm Tr}X^2={\rm const}\ne 0,
\end{equation}
assuming that $\mu,\,\nu={\rm const}$ and $\rho,\,\sigma$
are independent of $Y$, and using the
differential consequence of (2.2) such that
\begin{equation}
%<2.3>
{\rm Tr}X\ddot X = -{\rm Tr}{\dot X}^2,
\end{equation}
we find
\begin{equation}
%<2.4>
\sigma = 1,\qquad \rho = \mu\frac{{\rm Tr}{\dot X}^2}{{\rm
Tr}X^2}.
\end{equation}
Inserting (2.4) into (2.1), rescaling $t\to \sqrt{\mu}\,t$ and
setting $\frac{\nu}{\sqrt{\mu}}=\beta$ we finally obtain the following
Newton-like equation:
\begin{equation}
%<2.5>
\ddot X + \beta \dot X + \frac{{\rm Tr}{\dot X}^2}{{\rm
Tr}X^2 }X = e^{iX}Ye^{-iX} - Y,\qquad X(0)=X_0,\quad \dot
X(0)=\dot X_0.
\end{equation}

We now discuss the asymptotic behaviour of (2.5).  An
immediate consequence of (2.5) is
\begin{equation}
%<2.6>
{\rm Tr}X\ddot X + \beta {\rm Tr}X\dot X + {\rm Tr}
{\dot X}^2 = 0.
\end{equation}
Hence using
\begin{equation}
%<2.7>
{\rm Tr}X\ddot X = \frac{d}{dt}{\rm Tr}X\dot X - {\rm
Tr}{\dot X}^2,
\end{equation}
we arrive at the following equation:
\begin{equation}
%<2.8>
\frac{d^2}{dt^2}{\rm Tr}X^2 + \beta\frac{d}{dt}{\rm Tr}X^2 = 0,
\end{equation}
subject to the initial data
\begin{equation}
%<2.9>
\frac{d}{dt}{\rm Tr}X^2\big\vert_{t=0} = 2{\rm Tr}X_0\dot
X_0,\qquad {\rm Tr}X^2\big\vert_{t=0} = {\rm Tr}X_0^2.
\end{equation}
The solution of (2.8) is
\begin{equation}
%<2.10>
{\rm Tr}X^2 = 2{\rm Tr}X_0\dot
X_0\hbox{$\frac{1}{\beta}$}(1-e^{-\beta t}) + {\rm
Tr}X_0^2.
\end{equation}
It thus appears that whenever $\beta > 0$ then the solution
to (2.5) approaches the orbit such that
\begin{equation}
%<2.11>
{\rm Tr}X^2 = \hbox{$\frac{2}{\beta}$}{\rm Tr}X_0\dot
X_0 + {\rm Tr}X_0^2.
\end{equation}
In the case with $\beta=0$ the solution (2.10) takes the
form
\begin{equation}
%<2.12>
{\rm Tr}X^2 = 2{\rm Tr}X_0\dot X_0 t + {\rm Tr}X_0^2,
\end{equation}
i.e. there is no solution to (2.8) on the orbit unless
\begin{equation}
%<2.13>
{\rm Tr}X_0\dot X_0 = 0,
\end{equation}
that is the time evolution starts from the orbit which is
then by virtue of (2.12) and (2.13) the invariant set.
Clearly, the same holds true for $\beta < 0$.

Suppose that $\beta>0$ and ${\rm Tr}X^2\ge0$ for all
$X\in \hbox{\st\symbol{103}}$.  As we have shown above the
solution to (2.5) approaches then the orbit whose parameters
depend via (2.11) on the initial data.  Furthermore,
it is clear that (2.5) has the asymptotics
\begin{eqnarray}
%<2.14>
&&\ddot X + \beta \dot X + \frac{{\rm Tr}{\dot X}^2}{r^2
}X = \left(e^{iX}Ye^{-iX}\right)\big\vert_{{\rm Tr}X^2=r^2}
- Y,\\
&&{\rm Tr}X^2 = r^2,
\end{eqnarray}
where $r=\sqrt{\frac{2}{\beta}{\rm Tr}X_0\dot X_0+{\rm
Tr}X_0^2}$.  Evidently, (2.15) can be regarded as a
constraint.  In section (4.2) (see also appendix) we
demonstrate by an example of the $SU(2)$ group that the
solutions of the unconstrained equation (2.14) of the form
\begin{equation}
%<2.16>
\ddot X + \beta \dot X + \frac{{\rm Tr}{\dot X}^2}{r^2
}X = \left(e^{iX}Ye^{-iX}\right)\big\vert_{{\rm Tr}X^2=r^2}
- Y,
\end{equation}
where $r$ is a constant, approach the orbit ${\rm
Tr}X^2=r^2$ regardless of the initial data, that is this
orbit is a universal attracting set.  The only exception are
the initial conditions such that
\begin{equation}
%<2.17>
X_0 = \mu Y,\qquad \dot X_0 = \nu Y,\qquad \mu,\,\nu\in{\bf R}.
\end{equation}
These initial data correspond to the following ansatz for
the solution to (2.16):
\begin{equation}
%<2.18>
X(t) = x(t)Y,
\end{equation}
with the scalar $x(t)$, reducing (2.16) to
\begin{equation}
%<2.19>
\left[\ddot x + \beta \dot x + \frac{{\rm Tr}Y^2}{r^2
}x\dot x^2\right]Y =
\left[\left(e^{iX}Ye^{-iX}\right)\big\vert_{{\rm
Tr}X^2=r^2}\right]\big\vert_{X=xY} - Y.
\end{equation}
We note that (2.19) is not any asymptotics of the equation
\begin{equation}
%<2.20>
\ddot x + \beta \dot x + \frac{{\dot x}^2}{x} = 0,
\end{equation}
arising from inserting the ansatz (2.18) into (2.5).
\section{Periodic solutions to the Newton-like equation}
This section is devoted to the periodic solutions to the
abstract Newton-like equation (2.5).  Consider
eq.\ (2.5).  Our aim is to study the solutions to (2.5) on the
orbit, therefore we write (2.5) as
\begin{eqnarray}
%<3.1>
&&\ddot X + \beta \dot X + \frac{{\rm Tr}\dot X^2}{c}X =
e^{iX}Ye^{-iX} - Y,\\
&&{\rm Tr}X^2 = {\rm const} = c.
\end{eqnarray}
On using (3.1), (3.2) and the identity
\begin{equation}
%<3.3>
{\rm Tr}(e^{iX}Ye^{-iX} - Y)^2 = -2{\rm Tr}Y(e^{iX}Ye^{-iX} - Y),
\end{equation}
we arrive at the following relation:
\begin{equation}
%<3.4>
{\rm Tr}\ddot X^2 + 2\beta {\rm Tr}\dot X\ddot X +
\beta^2{\rm Tr}\dot X^2 - \frac{({\rm Tr}\dot X^2)^2}{c} =
-2{\rm Tr}Y(e^{iX}Ye^{-iX} - Y).
\end{equation}
Furthermore, an immediate consequence of (3.1) is
\begin{equation}
%<3.5>
{\rm Tr}Y\ddot X + \beta {\rm Tr}Y\dot X + \frac{{\rm
Tr}\dot X^2}{c}{\rm Tr}YX = {\rm Tr}Y(e^{iX}Ye^{-iX} - Y).
\end{equation}
Assuming that
\begin{equation}
%<3.6>
{\rm Tr}YX = {\rm const} = \kappa,
\end{equation}
we obtain from (3.5)
\begin{equation}
%<3.7>
\frac{\kappa}{c}{\rm Tr}\dot X^2 = {\rm Tr}Y(e^{iX}Ye^{-iX}
- Y).
\end{equation}
Inserting (3.7) into (3.4) we get
\begin{equation}
%<3.8>
{\rm Tr}\ddot X^2 + 2\beta {\rm Tr}\dot X\ddot X +
\left(\beta^2 + \frac{2\kappa}{c}\right){\rm Tr}\dot X^2 - \frac{1}{c}({\rm
Tr}\dot X^2)^2 = 0.
\end{equation}
We now define
\begin{equation}
%<3.9>
V = X - \frac{{\rm Tr}XY}{{\rm Tr}Y^2}Y,
\end{equation}
where $Y\ne 0$.  Evidently, if (3.6) holds then
\begin{equation}
%<3.10>
\dot V = \dot X,
\end{equation}
and (3.8) can be written as
\begin{equation}
%<3.11>
{\rm Tr}\ddot V^2 + 2\beta {\rm Tr}\dot V\ddot V +
\left(\beta^2 + \frac{2\kappa}{c}\right){\rm Tr}\dot V^2 - \frac{1}{c}({\rm
Tr}\dot V^2)^2 = 0.
\end{equation}
Further, it follows immediately from (3.9) that
\begin{equation}
%<3.12>
{\rm Tr}VY = 0,\qquad {\rm Tr}V^{\rm{(n)}}Y = 0,
\end{equation}
where $V^{\rm{(n)}}$ is the n-{\em th} time derivative of
$V$, and
\begin{equation}
%<3.13>
{\rm Tr}V^2 = {\rm Tr}X^2 - \frac{({\rm Tr}XY)^2}{{\rm
Tr}Y^2}.
\end{equation}
Clearly, whenever (3.2) and (3.6) take place then
\begin{equation}
%<3.14>
{\rm Tr}V^2 = {\rm const},\qquad {\rm Tr}V\dot V = 0.
\end{equation}
Suppose now (see (3.10)) that
\begin{equation}
%<3.15>
{\rm Tr}\dot V^2 = {\rm Tr}\dot X^2 = {\rm const} = \tau,
\end{equation}
and that the algebra $\hbox{\st\symbol{103}}$ is three
dimensional, then in view of (3.12), (3.14) and (3.15) the
triple $\{Y,\,V,\,\dot V\}$ forms the orthogonal basis of
$\hbox{\st\symbol{103}}$.  On expanding $\ddot V$ in this
basis:
\begin{equation}
%<3.16>
\ddot V = \mu V + \nu Y + \rho \dot V,
\end{equation}
and using a differential consequence of (3.15) such that
\begin{equation}
%<3.17>
{\rm Tr}\dot V \ddot V = 0,
\end{equation}
together with (3.12) we arrive at the following relation:
\begin{equation}
%<3.18>
\ddot V = \mu V.
\end{equation}
Hence using (3.14) one obtains
\begin{equation}
%<3.19>
{\rm Tr}V\ddot V = -{\rm Tr}\dot V^2 = \mu {\rm Tr}V^2.
\end{equation}
Therefore,
\begin{equation}
%<3.20>
\ddot V = -\frac{{\rm Tr}\dot V^2}{{\rm Tr}V^2}V,
\end{equation}
which leads to
\begin{equation}
%<3.21>
{\rm Tr}\ddot V^2 = \frac{({\rm Tr}\dot V^2)^2}{{\rm
Tr}V^2}.
\end{equation}
Finally, inserting (3.21) into (3.11) and taking into
account (3.17) (see also (3.2), (3.6), (3.13) and (3.15)) we
get
\begin{equation}
%<3.22>
\tau\left[\tau\left(\frac{1}{c-\frac{\kappa^2}{\varepsilon}} -
\frac{1}{c}\right) + \beta^2 + \frac{2\kappa}{c}\right] = 0,
\end{equation}
where $\varepsilon = {\rm Tr}Y^2$.  The validity of (3.22) is
the condition for the existence of the solution to (2.5)
satisfying (3.2), (3.6) and (3.15).  In section 4.3 we show
that in the case of the $SU(2)$ group such the solution
corresponds to the periodic motion on the sphere $S_2$, more
precisely, the uniform circular motion in a parallel.  The
counterpart of $X$ is then the position vector in ${\bf
R}^3$ and (3.20) is simply the normal acceleration.  Notice
that the case $\tau=0$ corresponds to the equilibrium
solutions to the equation (2.5) belonging to the orbit
defined by (3.2).  In fact, taking into account (3.7) and
(3.15) when $\tau=0$ as well as using (3.3) one finds
\begin{equation}
%<3.23>
X = \lambda Y.
\end{equation}
Hence, by virtue of (3.2) we get
\begin{equation}
%<3.24>
\lambda = \pm \sqrt{\frac{c}{\varepsilon}}.
\end{equation}
The formula (3.23) related to the ansatz (2.19) is also
obtained in the case $c-\frac{\kappa^2}{\varepsilon}=0$ which
means that
\begin{equation}
%<3.25>
{\rm Tr}X^2{\rm Tr}Y^2 = ({\rm Tr}XY)^2.
\end{equation}

We end this section with the remark concerning equation
(3.1), when $\beta=0$ and $Y=0$.  Clearly, we then have
\begin{equation}
%<3.26>
\ddot X = -\frac{{\rm Tr}\dot X^2}{{\rm Tr}X^2}X,
\end{equation}
where ${\rm Tr}X^2={\rm const}=c$.  Furthermore, an
immediate consequence of (3.2) and (3.25) is
\begin{equation}
%<3.27>
{\rm Tr}\dot X^2 = {\rm const}.
\end{equation}
In the light of the interpretation of the formula (3.21) in
the case of the $SU(2)$ group mentioned above it is
plausible that then (3.26) corresponds to the geodesic
equations on the sphere $S_2$.
Obviously, the solutions to geodesic equations
involve the periodic uniform motion in great circles.
\section{Dynamics on the {\bi SU}(2) group}
\subsection{Group-invariant dynamical system}
This section deals with the concrete realization of the
abstract Newton-like equation (2.5) in the case of the
$SU(2)$ group.  We first derive the nonlinear dynamical
system generated by (2.5) satisfied by the group parameters.
Consider eq.\ (2.5).  On taking into account the following
realization:
\begin{equation}
%<4.1>
{\bf J} = \hbox{$\frac{1}{2}$}\bs,
\end{equation}
where $\bs=(\sigma_1,\sigma_2,\sigma_3)$
and $\sigma_i$, $i=1$, 2, 3, are the Pauli matrices, of the
infinitesimal generators ${\rm J}_i$, $i=1$, 2, 3, of the
group $SO(3)$ which is locally isomorphic to $SU(2)$, we can
write the general element of the Lie algebra $X(t)$ as
\begin{equation}
%<4.2>
X(t) = {\bf
x}(t)\bd\frac{\bs}{2},
\end{equation}
where ${\bf x}(t)$: ${\bf R}\to{\bf R}^3$ and the dot
designates the inner product.  Analogously,
\begin{equation}
%<4.3>
Y = {\bf a}\bd\frac{\bs}{2},
\end{equation}
where {\bf a} is a constant vector of ${\bf R}^3$.  Now,
inserting (4.2) and (4.3) into (2.5) and using the
elementary properties of the Pauli matrices we arrive at the
following nonlinear system of second-order differential
equations:
\begin{eqnarray}
%<4.4>
&&\ddot{\bf x} + \beta\dot{\bf x} + \frac{\dot{\bf x}^2}{{\bf
x}^2}{\bf x} = (\cos|{\bf x}|-1){\bf a} + \frac{\sin|{\bf
x}|}{|{\bf x}|}{\bf a}\times{\bf x} + (1-\cos|{\bf
x}|)\frac{({\bf a}\bd{\bf x}){\bf x}}{{\bf x}^2},\\
&&{\bf x}(0) = {\bf x}_0,\quad \dot{\bf x}(0) = \dot{\bf x}_0,\nonumber
\end{eqnarray}
where ${\bf a}\times{\bf x}$ designates the vector product
of vectors {\bf a} and {\bf x}, $|{\bf x}|=\sqrt{{\bf x}^2}$
stands for the norm of the vector {\bf x} and $\dot{\bf
x}\equiv\frac{d{\bf x}}{dt}$.

Notice that the orbits given by (2.2) are the
two-dimensional spheres
\begin{equation}
%<4.5>
{\bf x}^2 = {\rm const}.
\end{equation}
The linear stability analysis shows that the
equilibrium $\overline{{\bf x}}={\bf a}$,
$\overline{\dot{\bf x}}={\bf 0}$ is unstable and
the equilibrium $\overline{{\bf x}}=-{\bf a}$,
$\overline{\dot{\bf x}}={\bf 0}$ is stable for sufficiently
large $\beta>0$.

We now discuss the asymptotic behaviour of the system (4.4).
On making use of the relation
\begin{equation}
%<4.6>
{\rm Tr}AB = \hbox{$\frac{1}{2}$}{\bf a}\bd{\bf b}
\end{equation}
for $A={\bf a}\bd\frac{\bs}{2}$ and $B={\bf
b}\bd\frac{\bs}{2}$,
and taking into account (4.2) and (4.3) we arrive at the
following realization of (2.10) in the case of the $SU(2)$
group:
\begin{equation}
%<4.7>
{\bf x}^2 = 2{\bf x}_0\bd\dot{\bf
x}_0\hbox{$1\over\beta$}(1-e^{-\beta t}) + {\bf x}_0^2.
\end{equation}
It can be easily checked that whenever the initial data satisfy
\begin{equation}
%<4.8>
2{\bf x}_0\bd\dot{\bf x}_0 + {\bf x}_0^2 > 0,
\end{equation}
and $\beta>0$, then the solution to (4.4) approaches the
orbit
\begin{equation}
%<4.9>
{\bf x}^2 = \hbox{$2\over\beta$}{\bf x}_0\bd\dot{\bf x}_0 +
{\bf x}_0^2.
\end{equation}
Furthermore, it is clear that for the initial conditions
such that
\begin{equation}
%<4.10>
{\bf x}_0\bd\dot{\bf x}_0 = 0,
\end{equation}
and arbitrary $\beta$ the orbit is an invariant set.

The remaining solutions do not approach any orbit.  Namely,
for $\beta>0$ and
\begin{equation}
%<4.11>
\hbox{$2\over\beta$}{\bf x}_0\bd\dot{\bf x}_0 + {\bf x}_0^2
= 0,\qquad {\bf x}_0\bd\dot{\bf x}_0\ne0,
\end{equation}
the solutions to (4.4) tend asymptotically to the singular
point ${\bf x}={\bf 0}$.  This point is approached after a
finite period of time
\begin{equation}
%<4.12>
t_* = -\frac{1}{\beta}{\rm ln}\left(1+\frac{\beta{\bf
x}_0^2}{2{\bf x}_0\bd\dot{\bf x}_0}\right),
\end{equation}
for the initial data satisfying the inequality
\begin{equation}
%<4.13>
2{\bf x}_0\bd\dot{\bf x}_0 + {\bf x}_0^2 < 0.
\end{equation}
Finally, if $\beta\le 0$ and ${\bf x}_0\bd\dot{\bf x}_0\ne0$
then the trajectories go to infinity.

Let us now focus our attention on the most interesting
case when $\beta>0$ and the initial data satisfy the
inequality (4.8) so that the solution to (4.4) approaches
the sphere (4.9). We set for simplicity ${\bf a}=(0,0,a_3)$.
Based on the observations of section 4.3 discussing asymptotic
dynamics on the orbit (see formula (4.44)) we find that
whenever the following condition holds:
\begin{equation}
%<4.14>
\beta^2\sqrt{2\beta^{-1}{\bf x}_0\bd\dot{\bf x}_0 + {\bf
x}_0^2} < |a_3|\left(1 + \cos\sqrt{2\beta^{-1}{\bf
x}_0\bd\dot{\bf x}_0 + {\bf x}_0^2}\,\right),
\end{equation}
where $\beta>0$ and
\begin{equation}
%<4.15>
a_3\frac{\sin\sqrt{2\beta^{-1}{\bf
x}_0\bd\dot{\bf x}_0 + {\bf x}_0^2}}{\sqrt{2\beta^{-1}{\bf
x}_0\bd\dot{\bf x}_0 + {\bf x}_0^2}} \ne 0,
\end{equation}
then the system (4.4) has the limit cycle given by (see
(4.43) and (4.45))
\begin{eqnarray}
%<4.16>
&&x_3 = - \frac{\beta^2(2\beta^{-1}{\bf x}_0\bd\dot{\bf
x}_0+{\bf x}_0^2)}{a_3\left(1+\cos\sqrt{2\beta^{-1}{\bf
x}_0\bd\dot{\bf x}_0+{\bf x}_0^2}\right)},\\
&&x_1^2 + x_2^2 = R^2,
\end{eqnarray}
where the radius of the limit cycle (the circle (4.17)) is
\begin{equation}
%<4.18>
R = \sqrt{\left(2\beta^{-1}{\bf x}_0\bd\dot{\bf x}_0+{\bf
x}_0^2\right)\left[1 - \frac{\beta^4(2\beta^{-1}{\bf
x}_0\bd\dot{\bf x}_0+{\bf x}_0^2)}{\left(1+\cos\sqrt{2
\beta^{-1}{\bf x}_0\bd\dot{\bf x}_0+{\bf x}_0^2}
\right)^2a_3^2}\right]}\,.
\end{equation}
On the other hand, if the following inequality is valid (see
(4.48)):
\begin{equation}
%<4.19>
\beta^2\sqrt{2\beta^{-1}{\bf x}_0\bd\dot{\bf x}_0 + {\bf
x}_0^2} > |a_3|\left(1 + \cos\sqrt{2\beta^{-1}{\bf
x}_0\bd\dot{\bf x}_0 + {\bf x}_0^2}\,\right),
\end{equation}
then the trajectories tend to the equilibrium point such
that
\begin{equation}
%<4.20>
\overline{x}_1=0,\quad \overline{x}_2=0,\quad \overline{x}_3
= -{\rm sgn}\,a_3\,\,\sqrt{2\beta^{-1}{\bf x}_0\bd\dot{\bf x}_0
+ {\bf x}_0^2},\quad \overline{\dot{\bf x}} = {\bf 0},
\end{equation}
where ${\rm sgn}\,x$ is the sign function.

It follows directly from (4.49) that the critical value of
the bifurcation parameter $\beta$ is the implicit function
of the initial data given by the equation
\begin{equation}
%<4.21>
\beta_c^2\sqrt{2\beta_c^{-1}{\bf x}_0\bd\dot{\bf x}_0 + {\bf
x}_0^2} = |a_3|\left(1 + \cos\sqrt{2\beta_c^{-1}{\bf
x}_0\bd\dot{\bf x}_0 + {\bf x}_0^2}\,\right).
\end{equation}
It should be noted that in view of (4.14)--(4.18) the system
(4.4) has an infinite number of limit cycles.  Evidently,
these limit cycles are unstable.
\subsection{Asymptotic dynamical system}
We now study the nonlinear dynamical system derived from
(2.17) for the $SU(2)$ group.  In view of (4.4) such system
is of the form
\begin{equation}
%<4.22>
\ddot{\bf x} + \beta\dot{\bf x} + \frac{\dot{\bf x}^2}{r^2
}{\bf x} = (\cos r-1){\bf a} + \frac{\sin r }{r}{\bf
a}\times{\bf x} + \frac{1-\cos r}{r^2}({\bf a}\bd{\bf x}){\bf
x}.
\end{equation}
As with (4.4) we find that the equilibrium $\overline{{\bf x}}
= \pm r\frac{{\bf a}}{|{\bf a}|}$, $\overline{\dot{\bf x}}
={\bf 0}$, with the plus sign is unstable and the
equilibrium with minus sign is stable for large enough
$\beta>0$.

We now discuss the $SU(2)$ realization of the equation
(2.19).  On setting
\begin{equation}
%<4.23>
{\bf x} = x{\bf a},
\end{equation}
we obtain from (4.22) the following equation:
\begin{equation}
%<4.24>
\ddot x + \beta\dot x + \frac{{\bf a}^2}{r^2}\dot x^2x =
(1-\cos r)\left(\frac{{\bf a}^2}{r^2}x^2 - 1\right).
\end{equation}
An easy inspection shows that the equilibrium $\overline{x}=r/|{\bf a}|$,
$\overline{\dot x} = 0$ to (4.24) is unstable and the
equilibrium $\overline{x}=-r/|{\bf a}|$, $\overline{\dot x} = 0$
is stable for $\beta>0$.
We now return to the asymptotic system (4.22).  In appendix we
show that whenever $\beta>0$, then the solutions to (4.22)
approach the orbit
\begin{equation}
%<4.25>
{\bf x}^2 = r^2
\end{equation}
for arbitrary initial data.  The only exception are the
initial conditions of the form
\begin{equation}
%<4.26>
{\bf x}_0 = \mu{\bf a},\qquad \dot{\bf x}_0 = \nu{\bf
a},\qquad \mu,\,\nu\in{\bf R},
\end{equation}
corresponding to the case of the equation (4.24).

Suppose that $\beta>0$.  As with the system (4.4) we set for
simplicity ${\bf a}=(0,0,a_3)$.  Suppose that the bifurcation
parameter $\beta$ satisfies the inequality
\begin{equation}
%<4.27>
\beta < \sqrt{\frac{|a_3|(1+\cos r)}{r}},
\end{equation}
where $a_3\frac{\sin r}{r}\ne0$.  Using observations of
section 4.3 we find that then the system (4.22) has the limit
cycle of the form (see (4.43) and (4.45))
\begin{eqnarray}
%<4.28>
&&x_3 = - \frac{\beta^2r^2}{a_3(1+\cos r)},\\
&&x_1^2 + x_2^2 = R^2,
\end{eqnarray}
where the radius of the limit cycle (the circle (4.29)) is
\begin{equation}
%<4.30>
R = r\sqrt{1 - \frac{\beta^4r^2}{(1+\cos r)^2a_3^2}}\,.
\end{equation}
Further, for $\beta$ such that
\begin{equation}
%<4.31>
\beta > \sqrt{\frac{|a_3|(1+\cos r)}{r}},
\end{equation}
the solutions to (4.22) approach the equilibrium point
\begin{equation}
%<4.32>
\overline{x}_1 = 0,\quad \overline{x}_2 = 0,\quad
\overline{x}_3 = -{\rm sgn}\,a_3\,\,r,\quad
\overline{\dot{\bf x}} = {\bf 0}.
\end{equation}
Evidently, the critical value of the bifurcation parameter $\beta$ is
given by
\begin{equation}
%<4.33>
\beta_c = \sqrt{\frac{|a_3|(1+\cos r)}{r}}.
\end{equation}
From the discussion completed in section 4.3 it
follows that the system (4.22) has the global Hopf
bifurcation at $\beta=\beta_c$.  Phase portraits from
numerical integration of the system (4.22) for $\beta>0$ are
illustrated in Figs.\ 1 and 2.\\[\baselineskip]
\centerline{\fbox{Fig. 1 and Fig. 2}}
\subsection{Dynamics on the sphere $S_2$}
We now discuss the dynamics on the orbit, i.e. the two-
dimensional sphere $S_2$ which has been already mentioned above to
be attracting or invariant set for the investigated
nonlinear dynamical systems for $\beta>0$.  In view of
(4.22) the corresponding system can be written as
\begin{subequations}
\begin{eqnarray}
%<4.34>
&&\ddot{\bf n} + \beta\dot{\bf n} + \frac{\dot{\bf n}^2}{r^2
}{\bf n} = (\cos r-1){\bf a} + \frac{\sin r }{r}{\bf
a}\times{\bf n} + \frac{1-\cos r}{r^2}({\bf a}\bd{\bf n}){\bf
n},\\
&&{\bf n}^2 = r^2.
\end{eqnarray}
\end{subequations}
Clearly, eq.\ (4.34b) is a constraint on (4.34a).  By
switching over to the spherical coordinates such that
\begin{equation}
%<4.35>
{\bf n} =
(r\sin\theta\cos\varphi,r\sin\theta\sin\varphi,r\cos\theta),
\end{equation}
and setting (without loose of generality)
\begin{equation}
%<4.36>
{\bf a} = (0,0,a_3),
\end{equation}
we obtain from (4.34) the following system:
\begin{subequations}
\begin{eqnarray}
%<4.37>
&&\ddot \theta + \beta\dot\theta - \sin\theta\cos\theta\,\dot
\varphi^2 - a_3\frac{1-\cos r}{r}\sin\theta = 0,\\
&&\ddot \varphi + \beta\dot\varphi + 2{\rm
ctg}\theta\,\dot\theta\dot \varphi - a_3\frac{\sin r}{r} = 0.
\end{eqnarray}
\end{subequations}
We note that the system (4.37) is Lagrangian one.  The
Lagrangian is
\begin{equation}
%<4.38>
L = \frac{1}{2}(\dot\theta^2 + \sin^2\theta\,\,\dot\varphi^2) -
a_3\frac{1-\cos r}{r}\cos\theta,
\end{equation}
and the system (4.37) can be written in the form of the
Lagrange equations
\begin{eqnarray}
%<4.39>
\frac{d}{dt}\frac{\partial L}{\partial\dot\theta} -
\frac{\partial L}{\partial\theta} &=& Q_{\theta},\nonumber\\
\frac{d}{dt}\frac{\partial L}{\partial\dot\varphi} -
\frac{\partial L}{\partial\varphi} &=& Q_{\varphi},
\end{eqnarray}
where the generalized nonpotential forces are
\begin{equation}
%<4.40>
Q_{\theta} = -\beta\dot\theta,\qquad Q_{\varphi} =
-\beta\sin^2\theta\,\,\dot\varphi + a_3\frac{\sin
r}{r}\sin^2\theta.
\end{equation}
It should also be noted that identifying the term
$a_3(1-\cos r)/r$ with the acceleration of free fall one can
regard the system described by (4.38) and (4.39) as a
generalized spherical pendulum.  The standard spherical
pendulum is easily seen to correspond to the particular case
of $\beta=0$ and  $r=\pi$.  The case $\beta=0$ and $r=2\pi$
or $\beta=0$ and $a_3=0$ refers to the free motion on the
sphere.

Consider the system (4.37).  Let us assume that
$a_3\frac{\sin r}{r}\ne0$.  On setting $\theta={\rm const}$,
we find $\dot\varphi=\omega={\rm const}$, so we then deal
with the uniform circular motion in a parallel.
Furthermore, we have
\begin{subequations}
\begin{eqnarray}
%<4.41>
-\omega^2 &=& a_3\frac{1-\cos r}{r\cos\theta},\\
\beta\omega &=& a_3\frac{\sin r}{r}.
\end{eqnarray}
\end{subequations}
By virtue of (4.41a) $\theta\in(\frac{\pi}{2},\pi]$ (``South
of the equator'') for $a_3>0$, and
$\theta\in[0,\frac{\pi}{2})$ (``North of the equator'') for
$a_3<0$.  The singular case $\theta=\frac{\pi}{2}$ can be
treated as the motion in the equator with infinite angular
velocity.  In view of (4.41b) such a case corresponds to the
limit $\beta=0$.

We remark that the abstract formula (3.22) providing the
criterion of periodic motion on the orbit takes the form
\begin{subequations}
\begin{equation}
%<4.42a>
\omega^2\cos^2\theta + \beta^2 + 2\frac{a_3\cos\theta}{r} =
0,
\end{equation}
where $\omega=\dot\varphi={\rm const}$ and $\theta={\rm
const}$.  On the other hand, the abstract relations (3.7)
and (3.6) taken together yield
\begin{equation}
%<4.42b>
\omega^2 = -a_3\frac{1-\cos r}{r\cos\theta},
\end{equation}
\end{subequations}
i.e. the formula (4.41a) is obtained.  It can be easily
checked that the system (4.41) is equivalent to the system
(4.42).

We now return to the system (4.41). Let us assume that
$\beta\ne0$ and $r\ne\pi$, so we exclude the case of the
standard spherical pendulum.  An immediate consequence of
(4.41) is the formula
\begin{equation}
%<4.43>
\cos\theta = -\frac{\beta^2r}{a_3(1+\cos r)}.
\end{equation}
Hence, recalling that the case of $\beta>0$ is investigated,
we find
\begin{equation}
%<4.44>
\beta \le \sqrt{\frac{|a_3|(1+\cos r)}{r}}.
\end{equation}
It should also be noted that in view of (4.43) the radius of
the circle where the motion takes place is given by
\begin{equation}
%<4.45>
R = r\sqrt{1 - \frac{\beta^4r^2}{(1+\cos r)^2a_3^2}}.
\end{equation}

We now come to the analysis of the asymptotic dynamics on
the sphere $S_2$.  Consider the system (4.37).  Integrating
(4.37b) yields
\begin{equation}
%<4.46>
\dot\varphi = \left(a_3\frac{\sin
r}{r}\int\limits_{0}^{t}\sin^2\theta\,\,e^{\beta\tau}d\tau +
\dot\varphi_0\sin^2\theta_0\right)\frac{e^{-\beta
t}}{\sin^2\theta}\,.
\end{equation}
We discuss the most interesting case of $\beta>0$.  Assuming that
$\sin\theta\not\to0$ we obtain from (4.46) the following
asymptotic relation:
\begin{equation}
%<4.47>
\dot\varphi = a_3\frac{\sin
r}{r}\lim_{t\to\infty}\frac{\int\limits_{0}^{t}\sin^2\theta\,
\,e^{\beta\tau}d\tau}{\sin^2\theta\,\,e^{\beta t}} =
a_3\frac{\sin r}{r}\frac{1}{2{\rm ctg}\theta\,\dot\theta +
\beta}\,.
\end{equation}
Suppose now that $a_3\frac{\sin r}{r}\ne0$.  Inserting
(4.47) into (4.37b) we find that at asymptotic times
$\ddot\varphi=0$, i.e. $\dot\varphi={\rm const}$, which in
view of (4.47) leads to $\theta={\rm const}$.  Clearly, the
case $\sin\theta\to0$ corresponds to approaching the
equilibrium point $\theta=\pi$ for $a_3>0$ or $\theta=0$ for
$a_3<0$.  Indeed, the equilibrium $\theta=0$
for $a_3>0$ and $\theta=\pi$ for $a_3<0$ have been already
mentioned in section 4.1 to be unstable.  Taking into account
(4.41) we see that whenever $\beta>0$ and $a_3\frac{\sin
r}{r}\ne0$ then for $\beta$ satisfying (4.44) the
trajectories approach the circle given by (4.43).  On the
other hand, whenever $\beta$ fulfils the inequality
\begin{equation}
%<4.48>
\beta > \sqrt{\frac{|a_3|(1+\cos r)}{r}},
\end{equation}
then the trajectories tend to the equilibrium point
$\theta=\pi$ for $a_3>0$ and $\theta=0$ for $a_3<0$.
Therefore, the critical value of the bifurcation parameter
$\beta$ is
\begin{equation}
%<4.49>
\beta_c = \sqrt{\frac{|a_3|(1+\cos r)}{r}}.
\end{equation}
Notice that for $\beta=\beta_c$ the radius of the circle
given by (4.45) equals zero, that is the periodic trajectory
(limit cycle) reduces to the equilibrium point $\theta=\pi$
($\theta=0$).  Furthermore, as we have shown above the limit
cycle given by (4.43) and the equilibrium point $\theta=\pi$
($\theta=0$) are universal attracting sets for
$0<\beta<\beta_c$ and $\beta>\beta_c$, respectively.  We
conclude that whenever $a_3\frac{\sin r}{r}\ne0$ and
$\beta>0$ then the system (4.37) has a global Hopf
bifurcation (the bifurcation of a limit cycle from an
equilibrium point) at $\beta=\beta_c$.

We remark that since $\theta={\rm const}$, therefore at
asymptotic times $Q_{\theta}=0$ (see (4.40)).  Furthermore,
by virtue of (4.40) and (4.41b) we get $Q_{\varphi}=0$.
Thus we find that the system (4.37) is asymptotically
Hamiltonian one.
\section{Conclusion}
In this work we have introduced an abstract second-order
Newton-like differential equation on a general Lie algebra
such that orbits of the Lie-group action are attracting set.
By passing to the group parameters corresponding to the
concrete Lie algebra the Newton-like equation generates the
nonlinear dynamical system referred to as the
group-invariant one with the attracting set coinciding with
the orbit.  In other words, a method has been
found in this paper for constructing nonlinear systems with
the presumed form of an attracting set.  We have
demonstrated that in the case of the $SU(2)$ group the
attracting manifold is the universal attracting set for the
system generated by the asymptotic version of the abstract
Newton-like equation.  It is suggested that such
universality holds true, at least in some regions of a phase
space, for general Lie groups.  In order to clarify the
asymptotic behaviour of the nonlinear dynamical systems
arising in the case with the $SU(2)$ group we have studied
the dynamical system on the sphere $S_2$ corresponding to
the generalized spherical pendulum.  Such system has been
shown to have the global Hopf bifurcation.  Based on this
observation we have found that the group-invariant dynamical
system is an interesting example of the system which has an
infinite number of unstable limit cycles.  On the other
hand, it has turned out that the asymptotic dynamical system
has a global Hopf bifurcation.  We note that the discussed limit
cycles (circles) arising in the case of the $SU(2)$ group
coincide with sections of the sphere $S_2$ by the planes
perpendicular to the vector {\bf a}.  In our opinion, such a
coincidence takes place in the general case of an orientable
two-dimensional Lie-group manifold, that is the concrete
form of the limit cycles (whenever they exist) is determined
by the orbit of the Lie-group action and the direction of
the element $Y$.  Thus we suggest that the actual treatment
provide a method for constructing nonlinear dynamical
systems with the presumed form of a limit cycle.  The 
example of the $SU(2)$ group discussed herein with simple 
oscillatory dynamics on the orbit is a model one.  We recall 
that numerous group-theoretic constructions are tested by 
the example of the $SU(2)$ group.  It seems that in the case 
of groups with higher-dimensional orbits the dynamics shoul 
be much more complex.  Since we do not know any criterion 
excluding symmetry of the chaotic attractor, therefore we 
should admit that the dynamics would be chaotic one.  We 
also observe that even in the case with the two-dimensional 
orbits some open questions arise when the orbit is not a 
connected set.  For example, it is not clear referring to
the experience with the $SU(2)$ group what is the behaviour 
of the system having the orbit coinciding with the torus 
when the section by the plane would consist of two circles.  
The corresponding dynamics should be rather nontrivial.  In the
light of the above comments it seems that the
group-theoretic approach introduced herein would be a useful
tool in the theory of nonlinear dynamical systems,
especially in the bifurcation theory, as well as in such
branches of applied mathematics as for example the control
theory.
\begin{ack}
The computer simulations were performed with ``Dynamics''
written by Yorke [8].  We would like to thank Professor
Tomasz Kapitaniak for the source code of the ``Dynamics''.
\end{ack}
\newpage
\appendix
\section{}
We now demonstrate that the solutions to the system (4.22)
with $\beta>0$ approach the sphere ${\bf x}^2=r^2$ for
arbitrary initial data except those ones satisfying the
condition (4.26) referring to the case of the equation
(4.24).  Consider the system (4.22).  We assume for
simplicity (without loose of generality) that ${\bf
a}=(0,0,a_3)$. By switching over to the spherical
coordinates such that
\begin{eqnarray}
%<a.1>
&&x_1 = \rho \sin\theta\cos\varphi,\nonumber \\
&&x_2 = \rho\sin\theta\sin\varphi,\nonumber\\
&&x_3 = \rho\cos\theta,
\end{eqnarray}
we arrive at the following system:
\begin{subequations}
\begin{eqnarray}
%<a.2a>
&&\ddot\rho + \beta\dot\rho +
\rho(\dot\theta^2+\sin^2\theta\,\,\dot\varphi^2)\left(\frac{
\rho^2}{r^2}-1\right) + \dot\rho^2\frac{\rho}{r^2} =
a_3(1-\cos r)\cos\theta\left(\frac{\rho^2}{r^2}-1\right),\\
&&\ddot\theta +\beta\dot\theta -
\sin\theta\cos\theta\,\dot\varphi^2 +
\frac{2\dot\rho}{\rho}\dot\theta -
a_3\frac{1-\cos r}{\rho}\sin\theta = 0,\\
&&\ddot\varphi + \beta\dot\varphi + 2{\rm
ctg}\theta\,\dot\theta\dot\varphi +\frac{2\dot\rho}{\rho}
\dot\varphi- a_3\frac{\sin r}{r} = 0.
\end{eqnarray}
\end{subequations}
Integration of (A.2c) yields
\begin{equation}
%<a.3>
\dot\varphi = \frac{\left(a_3\frac{\sin
r}{r}\int\limits_{0}^{t}\rho^2\sin^2\theta\,\,e^{\beta\tau}d
\tau +
\dot\varphi_0\rho_0^2\sin^2\theta_0\right)e^{-\beta t}}{\rho^2\sin^2
\theta}.
\end{equation}
Suppose that $\beta>0$ and $\sin\theta\not\to0$.  We also
assume that $a_3\frac{\sin r}{r}\ne0$.  Notice that this
inequality implies $\rho\not\to0$ since then ${\bf
x}\equiv0$ is not a solution of (4.22).  Proceeding
analogously as with eq.\ (4.46) we arrive at the following
asymptotic relation:
\begin{equation}
%<a.4>
\dot\varphi = a_3\frac{\sin
r}{r}\frac{1}{\frac{2\dot\rho}{\rho}+2{\rm
ctg}\theta\,\dot\theta+\beta}\,.
\end{equation}
In view of (A.2c) this leads to $\ddot\varphi=0$, i.e.
$\dot\varphi={\rm const}$.  Hence we find that at asymptotic
times
\begin{equation}
%<a.5>
\frac{\dot\rho}{\rho} + {\rm ctg}\theta\,\dot\theta = {\rm
const}.
\end{equation}
Integrating (A.5) we obtain
\begin{equation}
%<a.6>
\rho\sin\theta = Ce^{\mu t},
\end{equation}
where $C$ and $\mu$ are constant.  Suppose that $\mu>0$.  By
virtue of (A.6) this implies $\rho\to\infty$.  Hence, taking
into account (A.2a)--(A.2c) and differential consequences of
(A.6) we obtain after some calculation
\begin{equation}
%<a.7>
-\beta\mu - \frac{\dot\rho^2}{r^2} - \frac{{\rm
tg}^2\theta\,(\mu\rho-\dot\rho)^2}{r^2} -
\frac{\rho^2\sin^2\theta\,\,\dot\varphi^2}{r^2} +
a_3\frac{1-\cos r}{r^2}\rho\cos\theta = \mu^2.
\end{equation}
But
$\frac{\rho^2\sin^2\theta\,\,\dot\varphi^2}{r^2}\gg a_3\frac{1
-\cos r}{r^2}\rho\cos\theta$.  Therefore, the left-hand side
of (A.7) is negative and the right-hand side is positive.  This
contradiction shows that $\mu$ in (A.6) cannot be positive.
Let us assume now that $\mu<0$.  Then $\rho\to0$ and in view of (A.1) ${\bf
x}\to0$.  But ${\bf x}=0$ is not the solution to (4.22) when
$a_3\frac{\sin r}{r}\ne0$.  Thus it turns out that $\mu=0$
and the asymptotic formula (A.6) reduces to
\begin{equation}
%<a.8>
\rho\sin\theta = C.
\end{equation}
Now, eqs.\ (A.2a)--(A.2c) and (A.8) taken together yield
\begin{equation}
%<a.9>
-\frac{(\rho\dot\rho)^2}{\rho^2\cos^2\theta} +
(r^2-\rho^2\sin^2\theta)\dot\varphi^2 + a_3(1-\cos
r)\rho\cos\theta = 0.
\end{equation}
On integrating (A.9) written with the help of (A.8) in the form
independent of $\theta$, one can easily find that
$\dot\rho\ne0$ at asymptotic times, leads to contradiction.
Therefore $\dot\rho=0$, i.e. $\rho={\rm const}$.  Hence by
virtue of (A.8) we have
$\theta={\rm const}$.  Assuming that $\rho\ne r$, using
(A.2a) and (A.2b), and taking into account that $\rho\ne0$,
$\sin\theta\ne0$, $\dot\varphi^2\ne0$, we get
$\rho\dot\varphi^2=0$, a contradiction.  Thus we find
$\rho=r$.  Notice that an immediate consequence of this
observation and eq.\ (A.9) is the formula (4.41a) describing
the uniform circular motion on the sphere ${\bf x}^2=r^2$.
Clearly, the case $\sin\theta\to0$ corresponds to
approaching the stable equilibrium $\overline{{\bf x}}=(0,0,-{\rm
sgn}\,a_3\,\,r)$, $\overline{\dot{\bf x}}={\bf 0}$ of (4.22). 
We finally note that in the case of the
ansatz (4.23) implying eq.\ (4.24) we have $\sin\theta\equiv0$ in
(A.1) and the system (A.2) becomes singular one.

\newpage
\noindent {\large Legends to figures}\\[2\baselineskip]
\noindent Fig. 1.  The system (4.22) with $\beta=0.1$, $r=1$, ${\bf
a}=(0,0,1)$ and ${\bf x}_0=(1.4,1,1)$, $\dot{\bf x}_0=(0.1,0.1,0.1)$.  The
parameter $\beta$ satisfies (4.27).  The initial condition
fulfils ${\bf x}_0^2>r^2$.  The trajectory starts from the
point marked with the cross.  Left: the projection on the
$(x_1,x_2)$ plane.  Right: the projection on the $(x_1,x_3)$
plane.\\[\baselineskip]
\noindent Fig. 2.  The system (4.22), where $\beta$, $r$, ${\bf a}$,
are the same as in Fig.\ 1, and ${\bf x}_0=(0.1,0.1,0.1)$, $\dot{\bf
x}_0=(0.5,0.5,0.5)$.  The initial data satisfy ${\bf x}_0^2<r^2$.
Left: the projection on the $(x_1,x_2)$ plane.  Right: the
projection on the $(x_1,x_3)$ plane.
\end{document}